\documentclass[12pt]{article}
\usepackage{amsfonts}
\usepackage{amsmath}
\usepackage{amssymb}
\usepackage{mathptmx}
\usepackage{geometry}
\usepackage{graphicx}
\usepackage{hyperref}
\usepackage{cancel}
\usepackage{longtable}
\usepackage{textcomp}
\usepackage{tikz}
\usepackage{bm}
\usepackage{mathrsfs}
\usepackage{filecontents}
\usepackage{times}
\usepackage{epsfig}
\usepackage{xcolor}
\usepackage{slashed}
\usepackage{latexsym}
\usepackage{verbatim}
\usepackage{extarrows}
\usepackage{multirow}
\usepackage{rotating}
\usepackage{colortbl}
\usepackage{indentfirst}
\usepackage{tensor}

\definecolor{mygray}{gray}{.9}
\definecolor{intnull}{RGB}{213,229,255}
\usepackage{arydshln}
\usepackage{diagbox}

\usepackage{caption}
\usepackage{tikz}
\usetikzlibrary{arrows,shapes,chains}
\usepackage{graphicx, subfig}
\begin{document}
\renewcommand{\thefootnote}{\fnsymbol{footnote}}
\baselineskip=16pt
\pagenumbering{arabic}
\vspace{1.0cm}
\begin{center}
{\Large\sf Distinct thermodynamic and dynamic effects produced by scale factors in conformally related Einstein-power-Yang-Mills black holes}
\\[10pt]
\vspace{.5 cm}

{Yang Li\footnote{E-mail address: 2120190123@mail.nankai.edu.cn} and Yan-Gang Miao\footnote{Corresponding author. E-mail address: miaoyg@nankai.edu.cn}}
\vspace{3mm}

{School of Physics, Nankai University, Tianjin 300071, China}

\vspace{4.0ex}
\end{center}
\begin{center}
{\bf Abstract}
\end{center}

We study the thermodynamics and dynamics of high-dimensional Einstein-power-Yang-Mills black holes in conformal gravity. Specifically, we investigate a class of conformally related black holes whose metrics differ by a scale factor. We show that a suitable scale factor cures the geodesic incompleteness and the divergence of Kretschmann scalars at the center of black holes. In the aspect of thermodynamics, we analyse the Hawking temperature, the entropy, and the specific heat, and verify the existence of second-order phase transitions. We find that the thermodynamics of this class of conformally related black holes is independent of scale factors.
In the aspect of dynamics, we find that the quasinormal
modes of minimally coupled scalar field perturbations are dependent on scale factors. Quite interesting is that the behavior of quasinormal mode frequencies also supports the independence of scale factors for the second-order phase transitions.
Our results show that the scale factors produce distinct thermodynamic and dynamic effects in the conformally related Einstein-power-Yang-Mills black holes, which provides an interesting connection between thermodynamics and dynamics of black holes in conformal gravity.

\section{Introduction}
\label{sec:intr}

General Relativity (GR) is the most widely accepted theory of gravitation because it has been tested by numerous observations~\cite{Will}. Here we mention some recent experimental tests in support of GR from various aspects,
such as from the gravitational wave~\cite{GW}, the black hole imaging~\cite{image}, and the X-ray~\cite{X-ray}, etc.
Despite the convincible experimental supports, GR encounters multiple severe problems. First of all, the singularity at the center of black holes is inevitable, where the curvature is divergent and the current theory comes into failure. It is hard to believe that a physically complete theory would allow the existence of divergence. Secondly, GR encounters failure when it is directly applied into the area of cosmology.
GR was modified~\cite{Kip} to include a cosmological constant to explain the cosmic inflation. From
the perspective of modern views, the introduction of the cosmological constant term is merely a
compromised disposal in order to explain the cosmic inflation, however, it does not provide an
ultimate answer about the physical origin of this term. Meanwhile, it seems to be impossible
to explain the observed line speed of the stardust on the galactic arms within the frame of GR
because the gravity produced by the luminous matter in the galaxy cannot prevent the stardust
from escaping with such a high speed. Although many fancy hypotheses have been proposed to
explain these two difficulties, such as the dark energy (to explain the cosmic inflation) and the dark
matter (to explain the speed of galactic stardust)~\cite{DM}, a persuasive experimental confirmation of
these hypotheses is still beyond our reach. Thirdly, a satisfactory theory of quantum gravity has not been established. The essential problem is that GR is non-renormalizable in the perspective of quantum field theory (QFT), and many theories in which gravity is consistent with the frame of quantum theory are very complicated in formalism, {\em e.g.}, string theory~\cite{Polchinski}, thus they fail to make applicable predictions for physical observations.

The three problems mentioned above imply that GR is probably not the ultimate theory of gravity, which prompts us to consider other possible schemes, such as conformal gravity.
As early as 1918, Weyl found~\cite{Weyl} a theory of gravity different from Einstein's, in which not only the diffeomorphic invariance but also the conformal invariance were considered. In a system already endowed with the diffeomorphic invariance, such as GR, the conformal transformation is equivalent to the Weyl transformation. Thus, we identify the conformal transformation with the Weyl transformation, and sometimes refer to the Weyl factor as a scale factor. The conformal invariance is a strong constraint, under which the freedom of choice of actions is very limited.
The original conformal action of gravity was constructed~\cite{Weyl} to be square of the Weyl tensor in the 4-dimensional spacetimes. However, such an action has no conformal invariance in the dimensions other than 4 due to the specific scaling behavior of the Weyl tensor. Therefore, an alternative version of conformal gravity was given~\cite{andp} in which a massless scalar field was introduced. The invariance of all physical quantities under the conformal transformation implies that the conformal invariance can be regarded as a gauge symmetry of spacetimes. However, it is obvious that the universe we live in does not possess the conformal invariance, otherwise all the fundamental particles would be massless. It was suggested~\cite{1605, 1611} that the conformal symmetry was simultaneously broken in the early stage of the universe, similar to the simultaneous breaking of electroweak gauge symmetry in some sense, and all the equivalent solutions of conformal gravity became solutions of inequivalent gravitational theories, {\em e.g.}, the conformal action could reduce to the Einstein-Hilbert action.
Since there exists an extra degree of freedom for the choice of conformal transformations in conformal gravity,
one can remove the singularity of conformal gravity at the center of black holes. For instance, the initial Big Bang singularity of the Friedman-Robertson-Walker (FRW) model is actually non-singular if measured~\cite{1605} by the Weyl tensor. As a result, the FRW spacetimes are conformally equivalent to the flat spacetimes, {\em i.e.}, the appearance of singularity is simply caused by an improper choice of scale factors.
Moreover, the inflation of universe and the problem of galactic stardust could be well explained~\cite{2006} in conformal gravity without introducing dark energy and dark matter, respectively.
It was also suggested~\cite{2011} in string theory that the conformal symmetry is an important property of the ultimate quantum gravity theory since the non-renormalizability can be fixed in conformal gravity. Generally speaking, the conformal symmetry is important even necessary if one wants to overcome the defects of GR.

As is known, Yang-Mills (YM) fields have been proved to be an exception of the black hole (BH) no-hair theorem in both GR~\cite{Bizon} and conformal gravity~\cite{53,1411}. Furthermore, the achievement of Yang-Mills theory in particle physics is so remarkable that the standard model has been established, see, for example, the monograph~\cite{Peskin}. On the other hand, the theory of Yang-Mills fields coupled to Einstein's gravity, {\em i.e.}, the Einstein-Yang-Mills (EYM) theory has been developed~\cite{61,659}, where numerical and analytic BH solutions have been found in various dimensions.
The dynamical properties of high-dimensional EYM black holes have been investigated~\cite{2005} recently through the resonance behaviors of EYM BHs under scalar field perturbations, characterized by quasinormal modes (QNMs)~\cite{Berti,Kokkotas,Nollert,Konoplya QNM}. QNMs are eigenmodes of oscillation of a dissipative system and
thus they consist of a real part ($\mathrm{Re}\omega$) denoting the pure oscillation and an imaginary part ($\mathrm{Im}\omega$) denoting the time scale of damping. The behaviors of $\mathrm{Re}\omega$ and $\mathrm{Im}\omega$ represent BHs' dynamic responses to perturbations.
Based on the significance of conformal gravity mentioned in the above paragraph, it is warrant for us to combine the Einstein-power-Yang-Mills (EPYM) theory\footnote{It is an extension of EYM theory. For the details, see Sect. 2.} with conformal gravity, {\em i.e.}, to find the BH solutions of the conformally related EPYM theory, to analyze the thermodynamics and dynamics of these BHs, and investigate the relationship between thermodynamics and dynamics.

The outline of this paper is as follows. In Sect.~\ref{sec:CEPYMBH}, we briefly review the conformal gravity and EPYM theory, and then generalize the static spherically symmetric EPYM BH solution to its conformally related one, {\em i.e.}, the conformal-Einstein-power-Yang-Mills (CEPYM) BH solution. We calculate the Ricci scalars and Kretschmann scalars of the CEPYM BHs and
find that their divergence at the center of black holes can be removed by choosing a suitable conformal transformation. We also verify the geodesic completeness in the CEPYM BH spacetimes.
In Sect.~\ref{sec:thermo}, we focus on the thermodynamics of the CEPYM BHs by analyzing the Hawking temperature, the entropy, and the second-order phase transition. We then turn to the dynamics of the CEPYM BHs by computing the QNMs of minimally coupled scalar field
perturbations in terms of the sixth-order WKB method in Sect.~\ref{sec: QNM}. The reason why we restrict
our attention to the minimally coupled scalar filed is to connect the dynamics and thermodynamics of the CEPYM BHs. Finally, we present our conclusions in Sect.~\ref{sec: conc}.

\section{EPYM black hole in conformal gravity}
\label{sec:CEPYMBH}

In a gravitational system with a diffeomorphic invariance, the essence of conformal transformations is the Weyl transformation, which is simply the metric $g_{\mu\nu}$ being multiplied by a scale factor $\Omega^{2}(x)$,
\begin{equation}
\tilde{g}_{\mu\nu}=\Omega^{2}(x) g_{\mu\nu},\label{metrictrans}
\end{equation}
and the transformation of the determinant $\sqrt{-g}$ depends on the spacetime dimension $D$,
\begin{equation}
\sqrt{-\tilde{g}}=\Omega^{D}(x)\sqrt{-g}.\label{dettrans}
\end{equation}
The action of conformal gravity proposed by Weyl~\cite{Weyl} takes the form,
\begin{equation}
I_{\rm W}=\int \mathrm{d}^{D}x \,\sqrt{-g}\, C_{\mu\nu\sigma\rho}C^{\mu\nu\sigma\rho},\label{weylact}
\end{equation}
where two Weyl tensors are fully contracted with each other. Note that the original Weyl tensor is
$\tensor {C}{_\mu_\nu_\sigma^\rho}$,
which is invariant under the conformal transformation, Eq.~(\ref{metrictrans}). However, when index $\rho$ is lowered, the covariant tensor, $\tensor {C}{_\mu_\nu_\sigma_\rho}$, is no longer invariant,
\begin{equation}
\label{Weyldown}
\tensor {\tilde{C}}{_{\mu\nu\sigma\rho}}=\Omega^{2}(x){C}{_{\mu\nu\sigma\rho}},
\end{equation}
and neither is its contravariant tensor,
\begin{equation}
\label{Weylup}
\tensor {\tilde{C}}{^{\mu\nu\sigma\rho}}=\Omega^{-6}(x){C}{^{\mu\nu\sigma\rho}}.
\end{equation}
In the 4-dimensional spacetimes, $\sqrt{-g}$ gives the scale factor $\Omega^{4}(x)$, see Eq.~(\ref{dettrans}). Thus, the Weyl action, Eq.~(\ref{weylact}), is invariant under the conformal transformation. However, in dimensions other than 4, this action is not conformally invariant.

Here we focus on an alternative action of conformal gravity~\cite{andp},
\begin{equation}
I_{\rm C}=\frac{1}{2}\int \mathrm{d}^{D}x \,\sqrt{-{g}} \phi\left(\frac{1}{4}\frac{D-2}{D-1}{R} \phi-{\square} \phi\right), \label{confact}
\end{equation}
where $\phi$ is a massless scalar field, $R$ the Ricci scalar, and ${\square}\equiv g^{\mu\nu}\nabla_\mu\nabla_\nu$ the covariant d'Alembertian.
As shown in Ref.~\cite{andp},
Eq.~(\ref{confact}) is invariant under the conformal transformations of ${g}_{\mu\nu}$ and $\phi$ as follows,
\begin{equation}
{\tilde{g}}_{\mu\nu}=\Omega^{2}(x) {g}_{\mu\nu},\label{hatmetrictrans}
\end{equation}
\begin{equation}
{\tilde\phi}=\Omega^{\frac{2-D}{2}}(x)\,\phi.\label{phitrans}
\end{equation}
where $\Omega(x)$ is the scale factor which is an arbitrary smooth function and respects the asymptotic structure of spacetimes. In our following contexts, $g_{\mu\nu}$ will describe a EPYM black hole, the scale factor $\Omega^{2}(x)$
will properly be chosen due to the conformal invariance, and thus the metric ${\tilde{g}}_{\mu\nu}$ which describes a family of conformally related EPYM black holes will be determined by Eq.~(\ref{hatmetrictrans}).

\subsection{Metric of CEPYM black holes}
\label{sec: metric and dimless parameters}

We focus on $SO(D-1)$ Yang-Mills fields coupled with Einstein's gravity, {\em i.e.}, the analytic solutions of Einstein-Yang-Mills (EYM) black holes~\cite{659,1562,0212288,0212022} and of Einstein-power-Yang-Mills (EPYM) black holes~\cite{EGBPYM}.

In $SO(D-1)$ gauge theory, the Yang-Mills invariant has the form,
\begin{equation}
\label{YMLagrangian}
\mathcal{F}_{\rm YM}=\sum_{a=1}^{(D-1)(D-2)/2}F^{a}_{\mu\nu}F^{a\mu\nu},
\end{equation}
where $F_{\mu\nu}^{a}=\partial_{\mu}A_{\nu}^{a}-\partial_{\nu}A_{\mu}^{a}+\frac{1}{2\sigma}C_{bc}^{a}A_{\mu}^{b}A_{\nu}^{c}$ are the field strengths of $SO(D-1)$ Yang-Mills fields with the structure constants $C_{bc}^{a}$ and coupling constant $\sigma$, and $A_{\mu}^{a}$ are the gauge potentials, the Latin indices, $a, b, c, \dots=1, 2, \dots, (D-1)(D-2)/2$, represent the internal space of the gauge group, and the Greek indices, $\mu, \nu, \alpha, \beta, \dots=0, 1, \dots, D-1$, describe $D$-dimensional spacetimes.
The action of $SO(D-1)$ gauge theory, $I_{\rm YM}=-\frac{1}{2}\int \mathrm{d}^{D}x \,\sqrt{-g}\mathcal{F}_{\rm YM}$, is conformally invariant only in the 4-dimensional spacetimes.

In order to have a conformally invariant version in $D$ dimensions, a power-Yang-Mills (PYM) invariant, $\left(\mathcal{F}_{\rm YM}\right)^{q}$, should be introduced, whose action, $I_{\rm PYM}=-\frac{1}{2}\int \mathrm{d}^{D}x \,\sqrt{-g}\left(\mathcal{F}_{\rm YM}\right)^{q}$, has the conformal invariance when the power exponent $q$ satisfies $4q=D$.
According to Ref.~\cite{EGBPYM}, the EPYM action takes the form,
\begin{equation}
\label{EPYMaction}
I_{\rm EPYM}=\frac{1}{2}\int \mathrm{d}^{D}x \,\sqrt{-g}\left[R-\left(\mathcal{F}_{\rm YM}\right)^{q}\right],
\end{equation}
which reduces to the EYM action if $q=1$, and the solutions of Eq.~(\ref{EPYMaction}) are
\begin{equation}
f(r)=\begin{cases}
1-\frac{4M}{(D-2)r^{D-3}}-\frac{Q_1 }{r^{4q-2}}-\frac{\Lambda}{3}r^2, \qquad Q_1=\frac{((D-3)(D-2)Q^2)^{q}}{(D-2)(D-1-4q)}, & \qquad 4q\neq D-1;\\
 1-\frac{4M}{(D-2)r^{D-3}}-\frac{Q_2 \ln r}{r^{D-3}}-\frac{\Lambda}{3}r^2, \qquad Q_2=\frac{((D-3)(D-2)Q^2)^{(D-1)/4}}{D-2}, & \qquad 4q=D-1.
\end{cases}
\end{equation}
Here we adopt only the first branch with $4q \neq D-1$ because we shall construct our CEPYM theory which will possess  the conformal invariance when $4q=D$ and $\Lambda=0$. The static spherically symmetric black hole solution of Eq.~(\ref{EPYMaction}) under Wu-Yang ansatz~\cite{WY, PB} reads
\begin{equation}
\label{metric}
\mathrm{d}s^2=-f(r)\mathrm{d}t^{2}+\frac{1}{f(r)}\mathrm{d}r^{2}+r^{2}\mathrm{d}\Omega_{D-2},
\end{equation}
\begin{equation}
\label{lapsefunc}
f(r)=1-\frac{4M}{(D-2)r^{D-3}}-\frac{Q_{1}}{r^{D-2}},
\end{equation}
\begin{equation}
Q_1 \equiv \frac{((D-3)(D-2)Q^{2})^{q}}{(D-2)(D-1-4q)},\label{q1}
\end{equation}
where $Q$ is the only non-zero gauge charge of $SO(D-1)$ gauge group and $M$ the BH mass. Note that the EPYM action has no conformal invariance although its second part, see Eq.~(\ref{EPYMaction}), the PYM action has such an invariance when $4q=D$.

By combining Eq.~(\ref{confact}) and Eq.~(\ref{EPYMaction}), we generalize the EPYM theory to conformal gravity and write the CEPYM action,
\begin{equation}
\label{CEPYMaction}
I_{\rm CEPYM}=\frac{1}{2}\int \mathrm{d}^{D}x \,\sqrt{-{g}} \left(\frac{1}{4}\frac{D-2}{D-1}{R} \phi^{2}-\phi{\square} \phi-\left(\mathcal{F}_{\rm YM}\right)^{q}\right),
\end{equation}
which is invariant under the conformal transformations, Eqs.~(\ref{hatmetrictrans}) and (\ref{phitrans}), when $D=4q$. Note that the CEPYM action reduces to the EPYM action when the conformal (Weyl) symmetry is simultaneously broken with the specific choices of $\phi=2\sqrt\frac{D-1}{D-2}$ and $\Omega^{2}(x)=1$.

After the Weyl symmetry
breaking, different scale factors correspond to different conformally related spacetimes, which results
in different observational signatures. Here we take the scale factor suggested in Ref.~\cite{1611},
\begin{equation}
\label{Weylfactor}
{\cal S}(r)\equiv \Omega^{2}(r)=\left(1+\frac{L^{2}}{r^{2}}\right)^{2N},
\end{equation}
where $N \in \mathbb{N}^{+}$ is a free positive integer parameter called the scale exponent of conformal transformations and $L$ a length scale.
We can see that the scale factor reduces to unity when $L\ll r$. In this limit, the EPYM theory is recovered. Moreover, this scale factor was also chosen in Refs.~\cite{1705, Pisin} where a conformally related Schwarzschild BH was obtained and its QNMs of scalar, electromagnetic, and axial gravitational perturbations were computed. As a result, we obtain from Eqs.~(\ref{hatmetrictrans}) and (\ref{metric})-(\ref{q1}) the metric of a conformally related EPYM black hole, in short, the metric of a CEPYM black hole,
\begin{equation}
\label{conformmetric}
\mathrm{d}{\tilde s}^{2}={\cal S}(r)\left[
-f(r)\mathrm{d}t^{2}+\frac{1}{f(r)}\mathrm{d}r^{2}+r^{2}\mathrm{d}\Omega_{D-2}
\right].
\end{equation}
It is worth mentioning that Eq.~(\ref{conformmetric}) actually represents a family of CEPYM BH spacetimes which are related by the scale factor Eq.~(\ref{Weylfactor}). Next, we shall show that the Ricci scalars and Kretschmann scalars of the CEPYM BH spacetimes have no singularities everywhere by choosing a suitable scale exponent and verify that the geodesic completeness is guaranteed in the CEPYM BH spacetimes.

For the sake of convenience in the following discussions, we introduce the dimensionless rescaling of $r$, $L$, and $Q$,
\begin{equation}
\frac{r}{M^{1/(D-3)}} \rightarrow r, \qquad
\frac{L}{M^{1/(D-3)}} \rightarrow L,
\qquad
\frac{Q}{M^{2(D-2)/(D(D-3))}} \rightarrow \hat{Q}.
\label{dimensionless parameter}
\end{equation}
Note that we still use the same symbols for the radial coordinate and length scale but replace the charge by $\hat{Q}$ after rescaling, which will not cause confusion in the contexts below. The lapse function Eq.~(\ref{lapsefunc}) can be recast into the form consisting of dimensionless quantities,
\begin{equation}
f(r)=1-\frac{4}{(D-2)r^{D-3}}-\frac{\hat{Q_1}}{r^{D-2}},
\label{lapsefunc2}
\end{equation}
\begin{equation}
\hat{Q}_1 \equiv \frac{((D-3)(D-2)\hat{Q}^{2})^{q}}{(D-2)(D-1-4q)},
\label{q1_2}
\end{equation}
but the form of ${\cal S}(r)$, see Eq.~(\ref{Weylfactor}), maintains unchanged under
such a dimensionless rescaling.

\begin{figure}[htbp]
    \centering
        \includegraphics[width=0.45\textwidth]{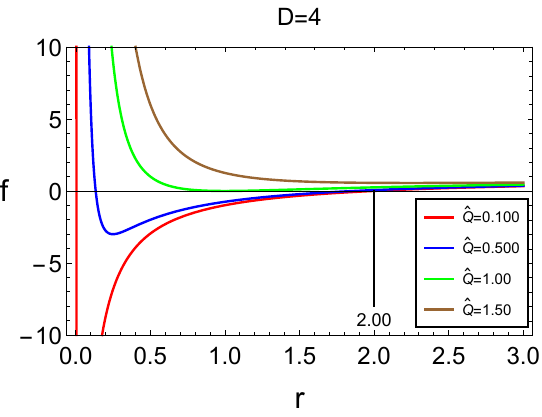}
        \includegraphics[width=0.45\textwidth]{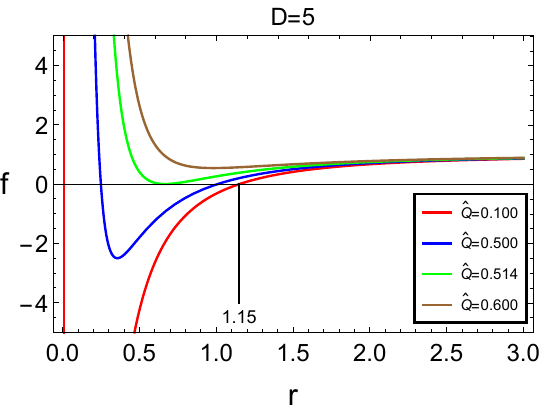}
        \includegraphics[width=0.45\textwidth]{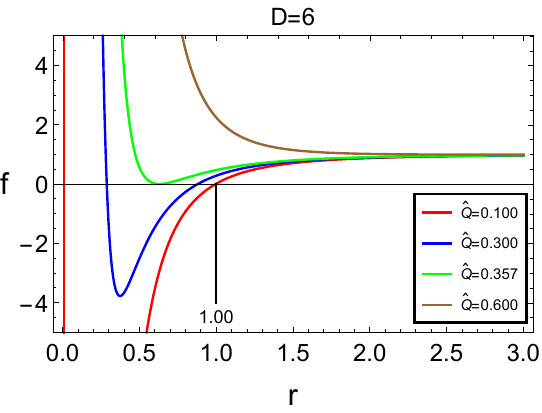}
	\caption{The lapse functions of 4, 5 and 6 dimensions with respect to the dimensionless radial coordinate. When $\hat{Q}= 0.100$, the horizon radius is $2.00$, $1.15$, and $1.00$ in 4, 5, and 6 dimensions,  respectively.}
\label{fig:lapse}
\end{figure}

We plot the image of Eq.~(\ref{lapsefunc2}) in Fig.~\ref{fig:lapse}, where the conditions that horizons exist can be seen clearly, {\em i.e.}, $\hat{Q} \lesssim 1.00$ in 4 dimensions, $\hat{Q} \lesssim 0.514$ in 5 dimensions, and $\hat{Q} \lesssim 0.357$ in 6 dimensions. In particular, these conditions also give the horizons of the extreme BHs when the dimensionless charge $\hat{Q}$ approximately takes $1.00$, $0.514$, and $0.357$, respectively. Moreover, the CEPYM BH has only one event horizon  when $\hat{Q}= 0.100$, where the horizon radius $r_{+}=2.00, 1.15, 1.00$ in 4, 5, and 6 dimensions,  respectively, and such a case will typically be chosen for analyzing the divergence or convergence for the Ricci scalars and Kretschmann scalars in Fig.~\ref{Rfig} and Fig.~\ref{Kfig}, respectively.

\subsection{Ricci and Kretschmann scalars of CEPYM black hole spacetimes}
\label{R and K}

The Ricci scalar is defined as the trace of Ricci tensors,
\begin{equation}
R\equiv g^{\mu\nu}R_{\mu\nu}.
\label{Ricci scalar}
\end{equation}
As shown in Ref.~\cite{EGBPYM}, the trace of the EPYM field equation vanishes, {\em i.e.}, $T^{a}{}_{a}=\frac{2-D}{D} R=-\frac{1}{2} (D-4q)\left({\cal F}_{YM}\right)^q=0$ when $4q=D$. Therefore, the Ricci scalar vanishes, namely, $R=0$ if $4q=D$. For the CEPYM black holes, when substituting Eqs.~(\ref{Weylfactor}), (\ref{conformmetric}), (\ref{lapsefunc2}) and (\ref{q1_2}) into Eq.~(\ref{Ricci scalar}),\footnote{In the contexts below, the condition $q=D/4$ will be considered without explanation each time.} we obtain the corresponding Ricci scalars to their first orders in the vicinity of $r=0$ in the 4-, 5-, and 6-dimensional spacetimes as follows,
\begin{eqnarray}
\tilde{R}_{4} &\approx& -\frac{(24 N^2+12N)\hat{Q}}{L^{4N}}r^{4N-4} ,
\nonumber \\
\tilde{R}_{5} &\approx& -\frac{ \left(96 \sqrt[4]{6} N^2+32 \sqrt[4]{6}N)\right)\hat{Q}^{5/4}}{L^{4N}}r^{4N-5} ,
\nonumber \\
\tilde{R}_{6} &\approx& -\frac{ \left(480 \sqrt{3} N^2+120 \sqrt{3}N \right)\hat{Q}^{3/2}}{L^{4N}}r^{4N-6}.
\end{eqnarray}
It is obvious that they are convergent in the limit of $r \rightarrow 0$ as long as $4N \geq D$.

We turn to the Kretschmann scalar defined as two fully contracted Riemann tensors,
\begin{equation}
\label{Kretschmann1}
K\equiv R^{\mu\nu\sigma\rho}R_{\mu\nu\sigma\rho}.
\end{equation}
For the EPYM black holes, we obtain the Kretschmann scalars to their first orders in the vicinity of $r=0$ in 4, 5, and 6 dimensions by substituting
Eqs.~(\ref{metric}), (\ref{lapsefunc2}) and (\ref{q1_2})  into Eq.~(\ref{Kretschmann1}),
\begin{eqnarray}
K_{4} &\approx& \frac{28 \hat{Q}^{2}}{r^{8}},\nonumber \\
K_{5} &\approx& \frac{420\sqrt{6}\hat{Q}^{5/2}}{r^{10}},\nonumber \\
K_{6} &\approx& \frac{29808\hat{Q}^{3}}{r^{12}},
\end{eqnarray}
which are singular at $r=0$. For the CEPYM black holes, when substituting Eqs.~(\ref{Weylfactor}), (\ref{conformmetric}), (\ref{lapsefunc2}) and (\ref{q1_2}) into Eq.~(\ref{Kretschmann1}), we obtain the corresponding Kretschmann scalars as follows,
\begin{eqnarray}
\tilde{K}_{4} &\approx& \frac{(28+40N+104N^{2}-64N^{3}+96N^{4})\hat{Q}^{2}}{L^{8N}}r^{2(4N-4)},\nonumber \\
\tilde{K}_{5} &\approx& \frac{(420\sqrt{6}+288\sqrt{6}N+1040\sqrt{6}N^{2}-576\sqrt{6}N^{3}+768\sqrt{6}N^{4})\hat{Q}^{5/2}}{L^{8N}}r^{2(4N-5)},\nonumber \\
\tilde{K}_{6} &\approx& \frac{(29808+12096N+56160N^{2}-27648N^{3}+34560N^{4})\hat{Q}^{3}}{L^{8N}}r^{2(4N-6)}.
\end{eqnarray}
Apparently, they are convergent in the limit of $r \rightarrow 0$ as long as $4N \geq D$. We notice that the constraint of $N$ for the CEPYM BHs is stronger than that given in Ref.~\cite{Pisin} for conformally related Schwarzschild BHs, which implies that the Kretschmann scalars in the CEPYM BH spacetimes will be divergent at the center of black holes if $N$ is not large enough in higher (than four) dimensions. Therefore, we conclude that $4N \geq D$ is the condition to remove the singularity of the Ricci and the Kretschmann scalars in the CEPYM BH spacetimes.

\begin{figure}[htbp]
    \centering
        \includegraphics[width=0.43\textwidth]{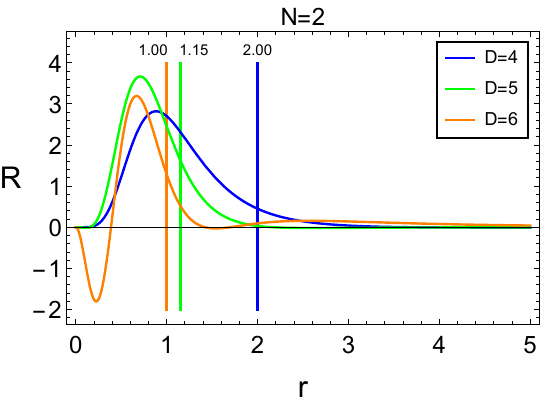}
        \includegraphics[width=0.45\textwidth]{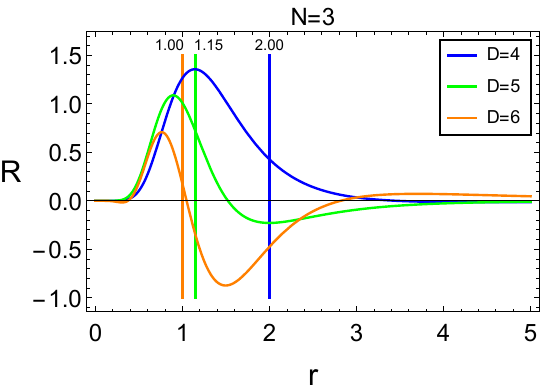}
        \includegraphics[width=0.45\textwidth]{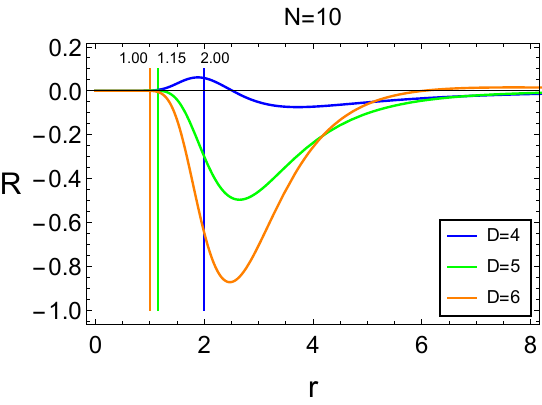}
	\caption{The Ricci scalar with respect to the dimensionless radial coordinate $r$, where $\hat{Q}=0.1$, $L=1$, and  $N=2, 3, 10$. The vertical lines mark the horizon positions in 4, 5, and 6 dimensions .}
\label{Rfig}
\end{figure}

In order to have a more intuitive description, we present the relationship between the Ricci scalar and the dimensionless radial coordinate $r$ in 4, 5, and 6 dimensions in Fig.~\ref{Rfig}, where the Ricci scalars in the EPYM BH spacetimes ($N=0$) are not shown because they apparently vanish in any dimensions as discussed above.

From Fig.~\ref{Rfig}, we can see that the Ricci scalars in the CEPYM BH spacetimes ($4N \geq D$) vanish at both $r=0$ and $r \rightarrow \infty$, which implies that the EPYM BHs and the CEPYM BHs are alike asymptotically. However, there is a major distinction between these two kinds of BHs in the near-horizon region. As we know, the Ricci scalar actually equals the trace of energy-momentum tensors. It is zero~\cite{EGBPYM} for the EPYM BHs since the Yang-Mills field is conformally coupled to gravity and the condition $4q=D$ is matched, but the situation is different for the CEPYM BHs, {\em i.e.}, the Ricci scalar can be positive and/or negative outside an event horizon. The reason is that the trace of energy-momentum tensors includes an extra contribution from the scalar field $\phi$ which is related to the scale factor via Eq.~(\ref{phitrans}). Although the Yang-Mills field still does not contribute to the trace, the scale factor is responsible for the non-vanishing trace of energy momentum tensors, and hence the non-vanishing Ricci scalar. Note that the null energy condition requires a positive trace of energy momentum tensors. In the regions where the Ricci scalar is negative, the null energy condition is violated and the gravitation becomes repulsive rather than attractive.

\begin{figure}[h]
    \centering
        \includegraphics[width=0.46\textwidth]{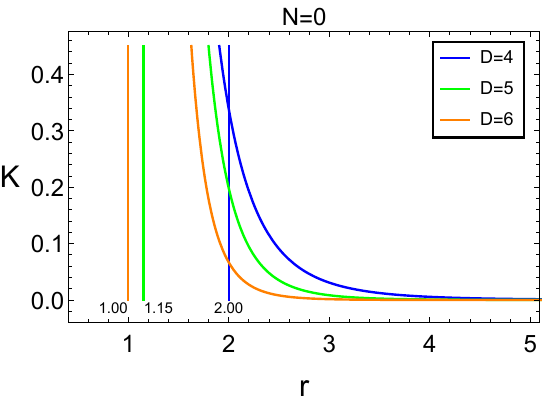}
    \vspace{5mm}
        \includegraphics[width=0.45\textwidth]{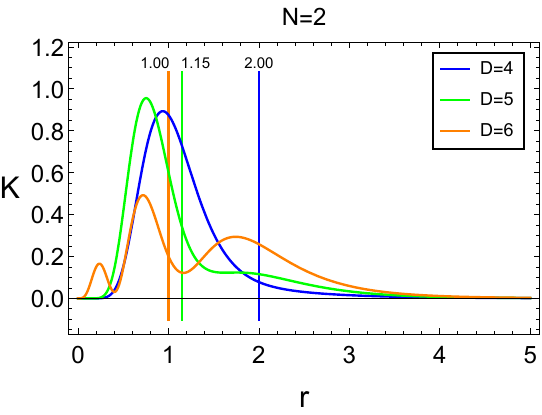}
    \vspace{5mm}
        \includegraphics[width=0.45\textwidth]{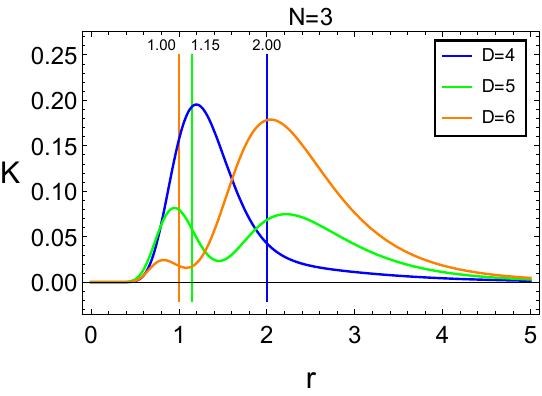}
    \vspace{5mm}
        \includegraphics[width=0.45\textwidth]{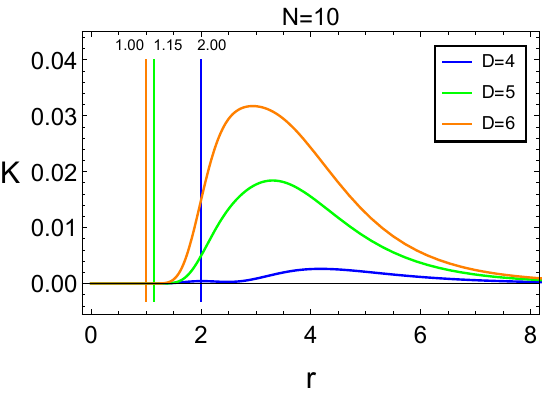}
	\caption{The Kretschmann scalar with respect to the dimensionless radial coordinate $r$, where $\hat{Q}=0.1$, $L=1$, and $N=0, 2, 3, 10$. Note that the case of $N=0$ correspond to the EPYM BH spacetimes, which is attached for comparison. The vertical lines mark the horizon positions in 4, 5, and 6 dimensions.}
\label{Kfig}
\end{figure}

Now we turn to the relationship between the Kretschmann scalar and the dimensionless radial coordinate $r$ in 4, 5, and 6 dimensions in Fig.~\ref{Kfig}.

From Fig.~\ref{Kfig}, we can see that the Kretschmann scalars in the EPYM BH spacetimes ($N=0$) are indeed singular at $r=0$, but they have no singularity in the CEPYM BH spacetimes ($4N \geq D$). In addition, the geometries of the EPYM BHs and CEPYM BHs are similar when $r \gg 1$ since the Kretschmann scalars of the two classes of BHs decrease when $r$ increases in a large $r$ region.

\subsection{Radial geodesic}
\label{sec:geodesics}
Now we prove the geodesic completeness in the CEPYM BH spacetimes by following Ref.~\cite{1605} and Ref.~\cite{Chakrabarty}. The equation that describes the radial geodesic motion takes the form,
\begin{equation}
g_{tt}\dot{t}^2+g_{rr}\dot{r}^{2}=-\delta_1,
\label{geodesic equation}
\end{equation}
where the dot represents the derivative of coordinates with respect to the affine parameter $\tau$, {\em e.g.}, $\dot{t}=\mathrm{d}t/\mathrm{d}\tau$, and $\delta_1$ equals zero for a null particle and one for a time-like particle. The energy is constant for any freely falling particles; here we restrict the attention to particles with only radial motion. The momentum related to the time component is as follows,
\begin{equation}
P_{t}=g_{tt} \dot{t}=-E.
\label{initial energy}
\end{equation}
Combining Eq.~(\ref{geodesic equation}) with Eq.~(\ref{initial energy}), we obtain the affine parameter in terms of  integration with respect to the radial coordinate $r$,
\begin{equation}
\label{affine parameter}
\Delta\tau= \int_{r_{f}}^{r_{i}} \mathrm{d}r \sqrt{-\frac{g_{tt}g_{rr}}{\delta_1 g_{tt} +E^2}},
\end{equation}
where $r_{i}$ represents the initial position of motion and $r_{f}$ the final position. When substituting Eq.~(\ref{conformmetric}) into Eq.~(\ref{affine parameter}), we derive the affine parameter in its convenient form in the CEPYM BH spacetimes,
\begin{equation}
\label{AP CEPYM}
\Delta\tau= \int_{r_{f}}^{r_{i}} \mathrm{d}r \sqrt{\frac{[{\cal S}(r)]^2}{E^2-\delta_1 {\cal S}(r)f(r)}}.
\end{equation}
In the following analyses, we shall adopt the dimensionless parameters introduced in Eq.~(\ref{dimensionless parameter}) for convenience.

\begin{figure}[htbp]
    \centering
        \includegraphics[width=0.5\textwidth]{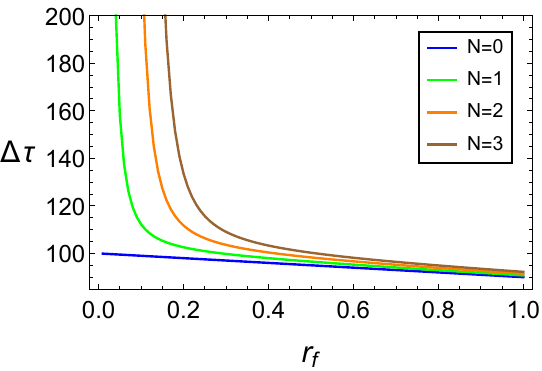}
	\caption{The affine parameter with respect to $r_{f}$, where $E=0.1$, $L=0.2$, and $r_{i}=10$.}
\label{fig:null}
\end{figure}

For null geodesics with $\delta_1=0$, the affine parameter is independent of the lapse function and spacetime dimension $D$. We plot the graph of the affine parameter with respect to $r_{f}$ in Fig.~\ref{fig:null}, where we fix the value of $r_{i}$ and vary the value of $r_{f}$.
In the EPYM BH spacetimes with $N=0$ and $\delta_1=0$, the integrand of Eq.~(\ref{AP CEPYM}) is a constant, so the affine parameter increases linearly as $r_{f}$ approaches the center of black holes, see the blue line in Fig.~\ref{fig:null}. In the CEPYM BH spacetimes, a particle moving towards the center of black holes along the radial null geodesics actually never reaches the center in a finite affine parameter, because the affine parameter is divergent when $r_{f}$ approaches the center, see, for instance, the cases of $N=1, 2, 3$. Here, the constraint $4N\geq D$ is not needed, since Eq. (\ref{AP CEPYM}) with $\delta_1=0$ does not include any information about spacetime dimensions. As a result, the radial null geodesics are incomplete in the EPYM BH spacetimes but complete in the CEPYM BH spacetimes.
Since the affine parameter of null geodesics is irrelevant to the lapse function in Eq.~(\ref{AP CEPYM}) due to $\delta_1=0$, a more general conclusion can be obtained: Null geodesics in the spacetimes described by Eq.~(\ref{Weylfactor}) and Eq.~(\ref{conformmetric}) with an arbitrary lapse function share the same behavior shown in Fig.~\ref{fig:null}.

For time-like geodesics with $\delta_1=1$, the result of integration Eq.~(\ref{AP CEPYM}) cannot be expressed analytically due to the complicacy of the integrand. Instead, we expand Eq.~(\ref{AP CEPYM}) in the vicinity of $r=0$ and analyze its approximate formulation. Starting with Eqs.~(\ref{geodesic equation}) and (\ref{initial energy}) and considering the dimensionless rescaling of Eq.~(\ref{dimensionless parameter}), we obtain the approximate formula assuming $r \ll 1$,
\begin{equation}
\dot{r}^2=\frac{\hat{Q}_1}{L^{4N}}r^{4N+2-D}+ \frac{E^2}{L^{8N}}r^{8N}.
\label{approx1}
\end{equation}

Now we discuss the competition of the two terms on the right-hand side of Eq.~(\ref{approx1}). There are three situations.

(a) When $4N+2-D > 8N$, the first term is infinitesimal compared to the second term and can be omitted. Note that $4N+2-D > 8N$ is equivalent to $D < 2-4N$, and that the scale exponent $N$ is a positive integer as well as the spacetime dimension $D$. In conclusion, the only allowed value of $N$ is zero, which means that the CEPYM BH spacetimes reduce to the EPYM ones. The spacetime dimension now satisfies $D < 2$, {\em i.e.}, $D = 0, 1$. In this case, the spacetime geometry is trivial since there is no gravitation in the dimensions lower than $2$.

(b) When $4N+2-D < 8N$, the second term on the right-hand side of Eq.~(\ref{approx1}) is infinitesimal compared to the first term and can be omitted. Now Eq.~(\ref{approx1}) leads to
\begin{equation}
\dot{r}=\frac{\sqrt{\hat{Q}_1}}{L^{2N}}r^{2N+1-D/2}.
\label{approx3}
\end{equation}
The constraint $4N \geq D$ obtained in Sect.~\ref{R and K} can be written as $4N+2-D \geq 2$. Combining with $4N+2-D < 8N$, we have $8N \mspace{2mu} > \mspace{2mu} 4N+2-D \mspace{2mu} \geq \mspace{2mu} 2$. For simplicity, we introduce a parameter defined as $a \equiv 2N+1-D/2$. Equivalently, we have $4N > a \geq 1$. The integration of Eq.~(\ref{approx3}) with respect to $r$ can be expressed as
\begin{equation}
\Delta\tau=\frac{L^{2N}}{\sqrt{\hat{Q}_1}} \int_{r_{f}}^{r_{i}} r^{-a} \mathrm{d}r.
\label{approx4}
\end{equation}
When the lower bound of $a$ is reached, {\em i.e.}, $a=1$ or $4N=D$, Eq.~(\ref{approx4}) leads to
\begin{equation}
\Delta\tau = \frac{L^{2N}}{\sqrt{\hat{Q}_1}} \left( \mathrm{ln} \mspace{3mu} r_{i} -\mathrm{ln} \mspace{3mu} r_{f} \right),
\label{approx5}
\end{equation}
which implies that the proper time is logarithmically divergent when $r_{f} \rightarrow 0$.
When $a>1$, Eq.~(\ref{approx4}) leads to
\begin{equation}
\Delta\tau = \frac{L^{2N}}{\sqrt{\hat{Q}_1}(a-1)} \left( \frac{1}{r_{f}^{a-1}} -\frac{1}{r_{i}^{a-1}} \right),
\label{approx6}
\end{equation}
which implies that the proper time is divergent in power law when $r_{f} \rightarrow 0$.

(c) When $4N+2-D = 8N$, the two terms on the right-hand side of Eq.(\ref{approx1}) have the same order and none of them can be omitted. The proper time becomes
\begin{equation}
\Delta\tau = \frac{L^{2N}}{(4N-1)}\left(\hat{Q}_1+\frac{E^2}{L^{4N}} \right)^{-1/2} \left( \frac{1}{r_{f}^{4N-1}} -\frac{1}{r_{i}^{4N-1}} \right).
\label{approx7}
\end{equation}
The condition under which Eq. (\ref{approx7}) diverges is $4N>1$. Because spacetime dimension $D$ is of course larger than $1$, we have $4N \geq D > 1$ for the CEPYM BH spacetimes, which gives rise to the completeness of
the radial time-like geodesics.

As a summary of this section, we have proven that the Ricci scalars and Kretschmann scalars are convergent at $r=0$, and both the null and time-like geodesics are complete in the CEPYM BH spacetimes with $4N \geq D$, which means that such CEPYM BH spacetimes have no singularity.

\section{Thermodynamics of CEPYM black holes}
\label{sec:thermo}

In this section, we study the Hawking temperature, the entropy, and the second-order phase transition of the CEPYM BHs.
The thermodynamics of conformally related black holes has been investigated~\cite{Bambi2017, Bambi2018}.
Here we focus on a CEPYM BH and find that it has a second-order phase transition, which is
probably the first example of conformally related black holes whose phase transition is revealed.

\subsection{Hawking temperature}
\label{sec:HT}

The discussion about the Hawking temperature is the preparation for us to investigate the
specific heat and second-order phase transition. In this way, we can make a comprehensive analysis
of thermodynamics for the CEPYM black holes.

The Hawking temperature is
\begin{equation}
\label{HTsurface}
T_{\rm H}=\frac{\kappa_{\rm H}}{2\pi},
\end{equation}
where $\kappa_{\rm H}$ is the surface gravity at an event horizon which is defined as
\begin{equation}
\label{surfacegravity}
\kappa_{\rm H}^{2}\equiv\left. -\frac{1}{2}\nabla_{\mu}\chi_{\nu}\nabla^{\mu}\chi^{\nu} \right|_{r=r_{+}}.
\end{equation}
In a $D$-dimensional static spacetime, $\chi_{\nu}$ is the time-like Killing vector, $\chi^{\nu}=(1,0,\ldots,0)$, and $r_{+}$ is horizon radius. For a CEPYM BH, the Hawking temperature takes the form,
\begin{equation}
\label{HTlapsefunc}
T_{\rm H}=\left. \frac{[{\cal S}(r)f(r)]^{\prime}}{4 \pi {\cal S}(r)} \right|_{r=r_{+}}= \left.\frac{f^{\prime}(r)}{4 \pi} \right|_{r=r_{+}},
\end{equation}
where the prime represents the derivative with respect to the radial coordinate $r$ and the definition of $r_{+}$, {\em i.e.}, $f(r_{+})=0$ has been considered.

Using Eq.~(\ref{lapsefunc2}), we rewrite Eq.~(\ref{HTlapsefunc}) in the dimensionless formalism,
\begin{equation}
T_{\rm H} = \frac{1}{4\pi}\left[\frac{4 (D-3)}{D-2} r_{+}^{2-D}-\left((D-3)(D-2) \hat{Q}^2\right)^{D/4} r_{+}^{1-D}\right],
\label{HT_explicit}
\end{equation}
where the horizon radii in the 4-, 5-, and 6-dimensional spacetimes have the following forms,
\begin{eqnarray}
r_+ &=& \sqrt{1-\hat{Q}^2}+1, \mspace{230mu}(D=4)
\nonumber\\
r_+ &=& \frac{1}{3}\left( A(\hat{Q})+\frac{4}{A(\hat{Q})}\right), \mspace{183mu}(D=5)
\nonumber\\
r_+ &=& \frac{1}{2}\left(\sqrt{B(\hat{Q})}+\sqrt{-B(\hat{Q})+\frac{2}{\sqrt{B(\hat{Q})}}} \right),\mspace{52mu}(D=6)
\label{radii}
\end{eqnarray}
with $A(\hat{Q})$ and $B(\hat{Q})$ defined by
\begin{eqnarray}
A(\hat{Q}) &\equiv& \sqrt[3]{\sqrt{729 \sqrt{6} \hat{Q}^5-64}-27 \sqrt[4]{6} \hat{Q}^{5/2}},
\nonumber\\
B(\hat{Q}) &\equiv& \frac{8 \sqrt[3]{2} \sqrt{3} \hat{Q}^3}{\sqrt[3]{\sqrt{1-6144 \sqrt{3} \hat{Q}^9}+1}}+\frac{\sqrt[3]{\sqrt{1-6144 \sqrt{3} \hat{Q}^9}+1}}{\sqrt[3]{2}}.
\label{functions in radii}
\end{eqnarray}
For the extreme BHs in 4, 5, and 6 dimensions, we have obtained the values of $\hat{Q}$ in Sect.~\ref{sec: metric and dimless parameters},  {\em i.e.}, $\hat{Q}_{\rm ext}\approx 1.00$, $0.514$, $0.357$, respectively. Substituting these values into Eqs.~(\ref{radii}) and (\ref{functions in radii}), we calculate the horizon radii of the three extreme BHs, $r_+^{\rm ext} \approx  1.00$, $0.667$, $0.630$, respectively.

\begin{figure}[htbp]
	\centering
        \includegraphics[width=0.5\textwidth]{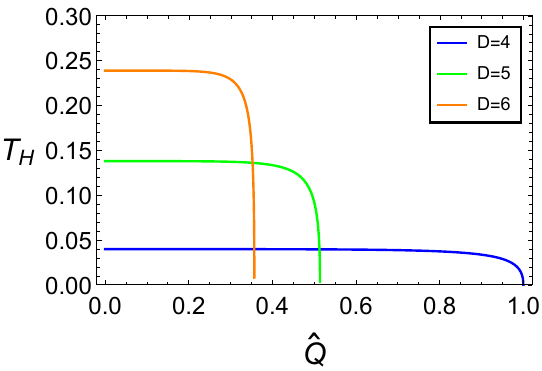}
	\caption{The Hawking temperature with respect to the dimensionless charge $\hat{Q}$, where $\hat{Q}_{\rm ext}$  approximately equals $1.00$, $0.514$, and $0.357$ for the 4-, 5-, and 6-dimensional extreme CEPYM BHs, respectively.}
\label{fig:HT}
\end{figure}

We plot the graph of the Hawking temperature with respect to the dimensionless charge for the 4-, 5-, and 6-dimensional CEPYM BHs in Fig.~\ref{fig:HT}. The Hawking temperature barely changes until near each extreme region, and then it rapidly diminishes and approaches zero. This phenomenon coincides with the Third Law of BH thermodynamics: The absolute zero that corresponds to the Hawking temperature of extreme BHs is unreachable due to the unreachable extreme configuration during the Hawking radiation.

\subsection{Entropy}
\label{sec:entropy}

We start with investigating the area of event horizons for the CEPYM BHs, which is different from that of the EPYM BHs because the scale factor imposes a non-trivial correction
to the area of event horizons,
\begin{equation}
\label{area}
A_{D-2}= \int \mathrm{d} \theta_{1} \ldots \mathrm{d} \theta_{D-2} \sqrt{h}=[{\cal S}(r_{+})]^{(D-2)/2}\,\frac{2 \pi^{(D-1)/2}r_{+}^{D-2}}{\Gamma \left(\frac{D-1}{2}\right)},
\end{equation}
where $A_{D-2}$ denotes the area of a $(D-2)$-dimensional spherical surface, $\theta_{i}$, $i=1, \ldots, D-2$, the $D-2$ angular coordinates, $h$ the determinant of the induced metric on the horizon $r_{+}$, and $\Gamma(x)$ the Gamma function.
In 4, 5 and 6 dimensions, $A_{D-2}$ yields,
\begin{eqnarray}
\label{area1}
A_2 &=& 4{\cal S}(r_{+}) \pi r_{+}^{2},\nonumber \\
A_3 &=& 2[{\cal S}(r_{+})]^{3/2}\pi^2 r_{+}^{3},\nonumber \\
A_4 &=& \frac{8 [{\cal S}(r_{+})]^{2} \pi^2 r_{+}^{4}}{3}.
\end{eqnarray}

\begin{figure}[h!]
	\centering
    \includegraphics[width=0.45\textwidth]{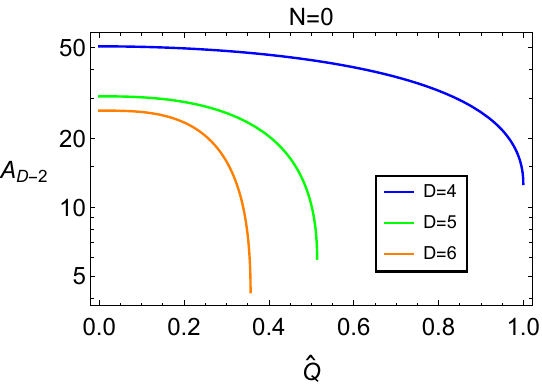}
		\vspace{5mm}
	\includegraphics[width=0.47\textwidth]{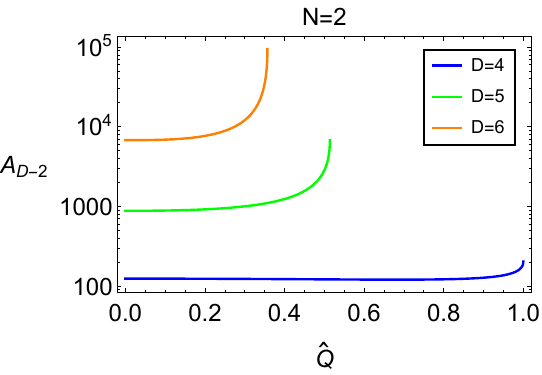}
		\vspace{5mm}
	\includegraphics[width=0.49\textwidth]{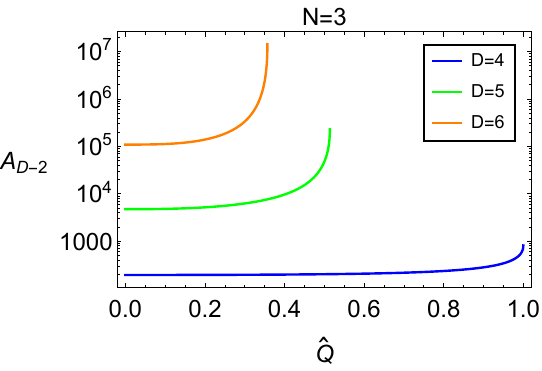}
		\vspace{5mm}
	\includegraphics[width=0.47\textwidth]{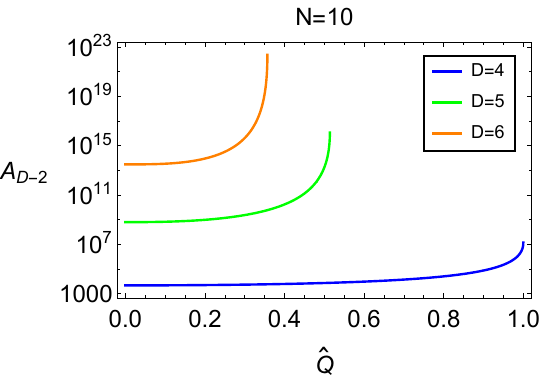}
	\caption{The  horizon area of the CEPYM BHs with respect to the dimensionless charge $\hat{Q}$ in 4, 5, and 6 dimensions for the cases of $N=2, 3, 10$, where the case of the EPYM BHs with $N=0$ is attached for comparison.}
\label{fig:Entropy}
\end{figure}

Substituting Eqs.~(\ref{Weylfactor}), (\ref{radii}) and (\ref{functions in radii}) into Eq.~(\ref{area1}), we plot the graph presenting the area of event horizons with respect to the dimensionless charge in 4, 5, and 6 dimensions in Fig.~\ref{fig:Entropy}, where the cases of $N=2, 3, 10$ are taken for the CEPYM BHs and the case of $N=0$ which corresponds to the EPYM BHs is attached for comparison.
Dimensionless parameter $L$ is set to be unity for convenience.
From this figure, we find that the area of event horizons of a CEPYM BH increases when the BH is approaching its extreme configuration but that of a EPYM BH decreases. The reason is that the scale factor has a non-trivial impact on the horizon area in the CEPYM BH spacetimes. In particular, the horizon area of a CEPYM BH is larger than that of a EPYM BH by several orders in magnitude. This blow-up of horizon areas is caused by the scale factor ${\cal S}(r)$ which equals a factor (larger than one) to the $(2N)$-th power, see Eq.~(\ref{Weylfactor}). As a result, the larger $N$ is, the higher orders in magnitude the area becomes.

The behavior of the horizon area with respect to the spacetime dimension in the CEPYM BH spacetimes is also different from that in the EPYM BH spacetimes. In the former case, the horizon area  in higher dimensions is larger, however, it is just the opposite in the latter case. Moreover, the disparities between the areas in two different dimensions increase with the increasing of the scale exponent in the CEPYM BH spacetimes. We thus conclude that the influence of ${\cal S}(r)$ on the horizon area is more obvious in higher dimensions.

Here a comment to $L$ is necessary. As a matter of fact, the theory does not provide any particular choices for the value of $L$. Theoretically, $L$ can be arbitrarily small such that ${\cal S}(r)$ approximates unity and the CEPYM BHs become arbitrarily close to the EPYM BHs. Indeed, a small $L$ would be more reasonable since Einstein's theory has been well verified by numerous experiments up to a decent precision. However, as mentioned above, $N$ is an arbitrary positive integer larger than $D/4$. Thus, the blow-up of horizon areas still happens even for a small $L$, as long as $N$ inflates to a very large value, {\em e.g.}, $N=100$. That is to say, this blow-up is inevitable, no matter how $L$ is sophisticatedly tuned. Nonetheless, such a blow-up implies that the area of event horizons does not represent the entropy of the CEPYM BHs in physics since the microscopic degrees of freedom of a thermodynamic system do not blow up under a spacetime dilation, which will be seen clearly in the First Law of BH thermodynamics below.

Next, we turn to discussions about the First Law of BH thermodynamics and derive the entropy of the CEPYM BHs. According to the first law of BH thermodynamics, the entropy should satisfy
\begin{equation}
\label{Smodified_overM}
\frac{\partial S_D}{\partial M}=\frac{1}{T_{\rm H}}.
\end{equation}
Note that the Hawking temperature does not depend on the scale factor ${\cal S}(r)$, see Eq.~(\ref{HTlapsefunc}), and the ADM mass $M$ maintains unchanged since ${\cal S}(r)$ does not influence the asymptotic flatness of the CEPYM BH spacetimes in the limit of $r \rightarrow \infty$. The entropy of a CEPYM BH is just the same as that of a EPYM BH, which means that the entropy itself is also independent of the scale factor ${\cal S}(r)$. Therefore, for both the EPYM BHs and the CPEYM BHs, we have their entropy,
\begin{equation}
S_D\equiv \frac{\pi^{(D-1)/2}r_{+}^{D-2}}{2\Gamma \left(\frac{D-1}{2}\right)}.
\end{equation}

\subsection{Specific heat and second-order phase transition}
\label{sec:2nd PT}

The Davies point at which a phase transition happens corresponds to the solution of the algebraic equation, $1/C_{Q}=0$, where the specific heat is defined as
\begin{equation}
\label{SpecificHeat1}
C_{Q}\equiv\left( \frac{\partial M}{\partial T}\right)_{Q}=1 \biggl{/} \left( \frac{\partial T}{\partial M}\right)_{Q}.
\end{equation}
As analyzed in the above subsection, the specific heat is independent of the scale factor ${\cal S}(r)$.
Note that the specific heat cannot be rescaled in terms of Eq.~(\ref{dimensionless parameter}) because $M$ should be regarded as a variable.

\begin{figure}[h!]
	\centering
    \includegraphics[width=0.5\textwidth]{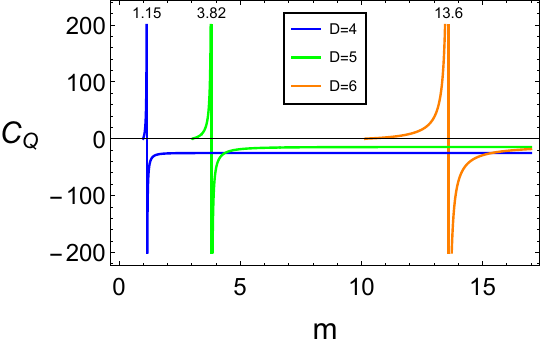}
	\caption{The dimensionless specific heat with respect to the dimensionless mass for the CEPYM BHs in 4, 5, and 6 dimensions.}
\label{fig:SpecificHeat}
\end{figure}

We adopt an alternative dimensionless rescaling here. At first, we derive the specific heat by using the original lapse function Eq.~(\ref{lapsefunc}) and the definition of the Hawking temperature Eq.~(\ref{HTlapsefunc}). Then, we rescale the specific heat in the dimensionless formalism in the $D$-dimensional spacetimes,
\begin{equation}
\frac{C_Q}{M^{(D-2)/(D-3)}} \longrightarrow  C_{Q}.
\label{Cq_rescaling_D_dim}
\end{equation}
With this rescaling, the dimensionless $C_{Q}$ becomes a function of the dimensionless mass $m$ defined as
\begin{equation}
m \equiv M/Q^{D(D-3)/(2D-4)}.
\end{equation}
Specifically, in 4 dimensions, Eq.~(\ref{Cq_rescaling_D_dim}) yields,
\begin{equation}
\frac{C_Q}{M^2} \longrightarrow C_{Q} = -\frac{2 \pi  \sqrt{1-m^{-2}} \left(\sqrt{1-m^{-2}}+1\right)^3}{ \sqrt{1-m^{-2}}-2 m^{-2}+1},
\label{Cq4D}
\end{equation}
where  $m=M/Q$.
In 5 and 6 dimensions, the rescaling of the specific heat takes the forms,
\begin{eqnarray}
\frac{C_Q}{M^{3/2}} & \longrightarrow & C_{Q}, \qquad (D=5)
\nonumber \\
\frac{C_Q}{M^{4/3}} & \longrightarrow & C_{Q}, \qquad (D=6)
\label{Cq5&6D}
\end{eqnarray}
where the corresponding $m$ equals $M/Q^{5/3}$ and $M/Q^{9/4}$, respectively. The derivation of the explicit forms of the specific heat in Eq. (\ref{Cq5&6D}) is tedious but straightforward, so we do not present it here.
Instead, we plot the dimensionless $C_{Q}$ in 4, 5, and 6 dimensions with respect to the dimensionless mass in Fig.~\ref{fig:SpecificHeat}.

From Fig.~\ref{fig:SpecificHeat}, we can see that the second-order phase transition exists in 4, 5, or 6 dimensions. The curves of specific heat for each spacetime are divided into two branches by a Davies point. By substituting Eq.~(\ref{Cq4D}) and Eq.~(\ref{Cq5&6D}) into $1/C_{Q}=0$ and solving it, we obtain the Davies points of the CEPYM BHs, $m_{\rm Davies}=2/\sqrt{3} \approx 1.15$, $m_{\rm Davies} \approx 3.82$, and $m_{\rm Davies} \approx 13.6$ in 4, 5, and 6 dimensions, respectively. In general, the Davies point corresponds to a larger value of $m$ in a higher dimension.
At the left branch in 4, 5, or 6 dimensions, the specific heat is positive before the
mass reaches the Davies point at which the second-order phase transition occurs, and the Hawking temperature increases with the increasing of $m$. After the mass exceeds the Davies point, i.e. at the right branch, the specific heat becomes negative, which implies that the Hawking temperature decreases with the increasing of $m$, and the lost energy converts into the Hawking radiation.
Moreover, $C_{Q}$ has a zero point at its left branch located at $m_{\rm ext}=1.00$ in 4 dimensions, $m_{\rm ext} \approx 3.00$ in 5 dimensions, and $m_{\rm ext} \approx 10.0$ in 6 dimensions, respectively, which corresponds to the condition of extreme BH configurations. The Hawking temperature vanishes in the extreme configurations, resulting in no Hawking radiation. Finally,
we can see that the Davies points locate near the extreme BH configurations when comparing $m_{\rm Davies}$ with $m_{\rm ext}$, which means that the second-order phase transition occurs near the extreme configurations.

In summary, we conclude that the thermodynamics the CEPYM BHs is independent of the scale factor.

\section{QNM of scalar field perturbations}
\label{sec: QNM}

The Klein-Gordon equation for a massless scalar field in a curved spacetime is
\begin{equation}
\label{KGeq1}
\nabla^{\mu}\nabla_{\mu}\Phi =0,
\end{equation}
or it can be written as
\begin{equation}
\label{KGeq2}
\frac{1}{\sqrt{-g}}\partial_{\mu}\sqrt{-g}g^{\mu\nu}\partial_{\nu}\Phi=0.
\end{equation}
In Ref.~\cite{1705}, the decomposition of scalar field $\Phi$ in the background of a 4-dimensional conformal non-singular BH was introduced. Here we generalize it to the $D$-dimensional spacetimes as follows,
\begin{equation}
\label{decompose}
\Phi=\sum_{l,m} \frac{1}{r^{(D-2)/2} [{\cal S}(r)]^{(D-2)/4}}e^{-i \omega t}\psi_{l}(r)\mathrm{Y}_{lm}
(\theta_{1}, \ldots, \theta_{D-2}),
\end{equation}
where $\mathrm{Y}_{lm}(\theta_{1}, \ldots, \theta_{D-2})$ stands for the spherical harmonics of $D-2$ angular coordinates, $0\leq \left( \theta_{1}, \theta_{2}, \ldots, \right.$ $\left. \theta_{D-3} \right) \leq \pi $, and $0 \leq \theta_{D-2} \leq 2 \pi$. Substituting Eq.~(\ref{decompose}) into Eq.~(\ref{KGeq2}), we obtain the radial equation,
\begin{equation}
\label{mastereq1}
f^{2}\psi_{l}^{\prime\prime}+ff^{\prime}\psi_{l}^{\prime}+\left( \omega^{2}-V_{\rm eff} \right)\psi_{l}=0,
\end{equation}
where $V_{\rm eff}$ is the effective potential,
\begin{equation}
\label{Veff}
V_{\rm eff}=f(r)\left[ \frac{l(l+D-3)}{r^2} + \frac{\left( f(r)\left(r^{(D-2)/2}[{\cal S}(r)]^{(D-2)/4}\right)^{\prime} \right)^{\prime}}{[{\cal S}(r)]^{(D-2)/4}r^{(D-2)/2}} \right].
\end{equation}
The prime represents the derivative with respect to $r$. Toshmatov et al. derived~\cite{1705} the master equation and the effective potential of the 4-dimensional minimally coupled scalar field. Our progress is to generalize their results to the D-dimensional spacetimes, see Eqs.~(\ref{mastereq1}) and (\ref{Veff}).
Using Eqs.~(\ref{Weylfactor}), (\ref{lapsefunc2}), (\ref{q1_2}), and (\ref{Veff}), we write the effective potential in its explicit form,
\begin{eqnarray}
V_{\rm eff} &=  & \frac{1 }{4 (D-2)r^{2 (D-1)}}\left[(D+2 l-4) (D+2 l-2) r^{D-2}+D(D-2) \hat{Q}_1+4 (D-2) r\right]
\nonumber\\
& &\times  \left[-(D-2) \hat{Q}_1+(D-2) r^{D-2}-4 r\right].
\end{eqnarray}
Again using the ``tortoise" coordinate $r_{*}$ defined as $\mathrm{d} r_{*}\equiv\mathrm{d}r / f(r)$, we finally derive the Schr$\ddot{\mathrm{o}}$dinger-like equation of $\psi_{l}(r)$,
\begin{equation}
\label{mastereq2}
\partial_{r_{*}}^2 \psi_{l}+\left( \omega^{2}-V_{\rm eff} \right)\psi_{l}=0.
\end{equation}

We use the WKB method to compute the quasinormal mode frequencies of Eq.~(\ref{mastereq2}). The WKB method is an approximation method for us to find the eigenvalues of Schr$\ddot{\mathrm{o}}$dinger-like equations, especially when the effective potential is too complicated to allow any analytic solutions. The first-order WKB method was introduced into BH perturbation theory by Schutz and Will~\cite{1st}, the third-order WKB by Iyer and Will~\cite{3rd}, and the sixth-order WKB by Konoplya~\cite{6th}. In the perspective of scattering theory, perturbations in a BH background can be regarded as waves scattered by effective potential $V_{\rm eff}$. The scattering is restricted by the boundary condition, which requires that the waves are purely ingoing at horizon and outgoing at infinity. For the scalar perturbation above, the boundary condition is
\begin{equation}
\label{Bcondition}
\psi_{l} \sim e^{\pm i \omega r_{*}}, \qquad r_{*} \rightarrow \pm \infty.
\end{equation}
Impacted by the boundary condition, the frequency $\omega$ is discretized. Because the perturbative scalar field is dissipated due to the existence of event horizons, $\omega$ is complex. These discrete frequencies are the quasinormal mode frequencies of perturbation, in short, quasinormal modes (QNMs). 

\begin{figure}[htbp]
    \centering
        \includegraphics[width=0.49\textwidth]{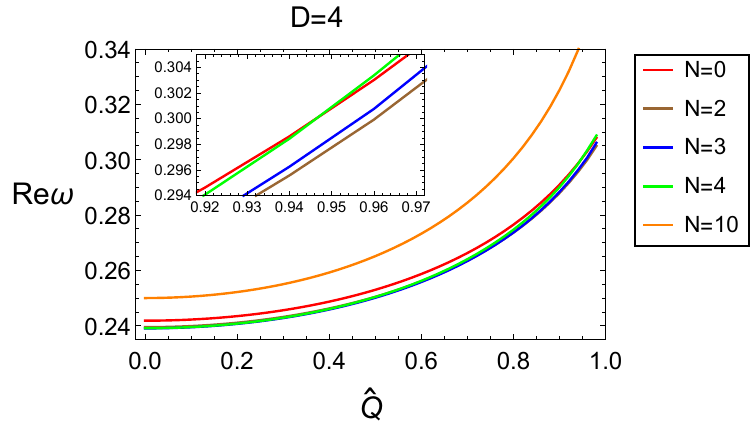}
        \includegraphics[width=0.49\textwidth]{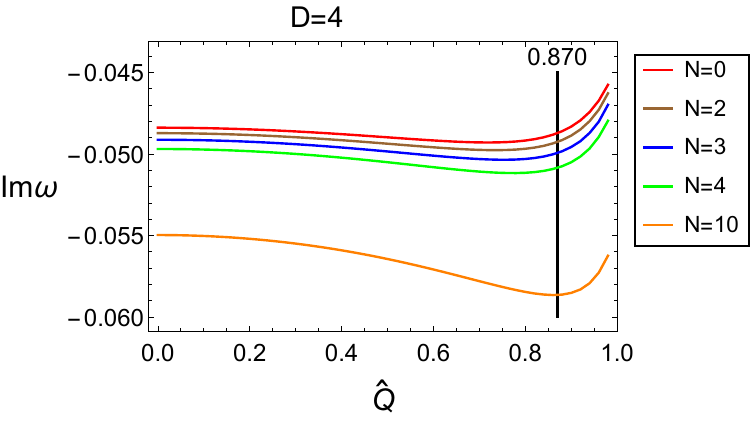}
        \vspace{5mm}
        \includegraphics[width=0.49\textwidth]{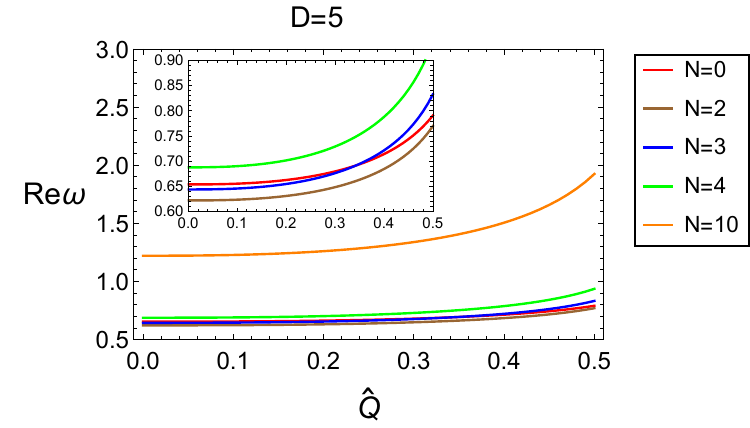}
        \includegraphics[width=0.49\textwidth]{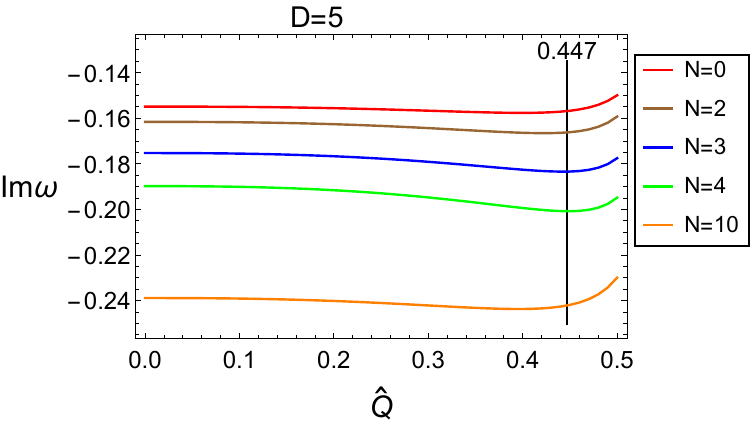}
        \vspace{5mm}
        \includegraphics[width=0.49\textwidth]{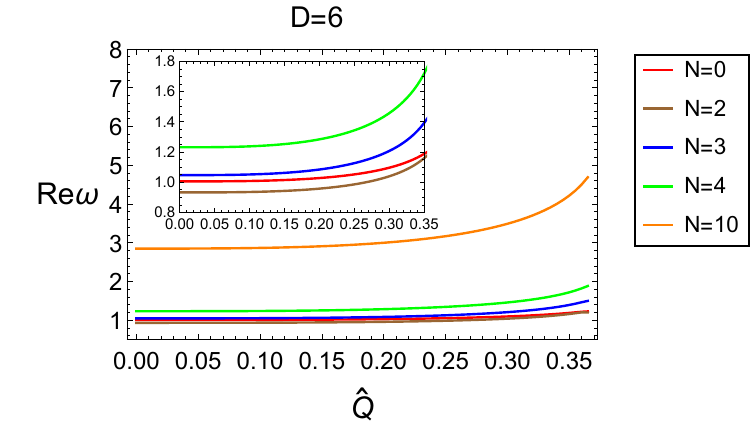}
    	\includegraphics[width=0.49\textwidth]{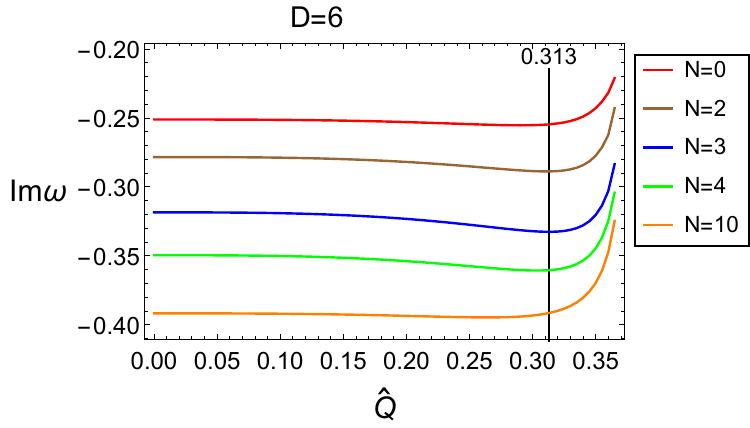}
	\caption{The real and imaginary parts of fundamental QNMs with respect to dimensionless charge $\hat{Q}$ for  different values of the scale exponent, $N=0, 2, 3, 4, 10$. For $N=0, 2, 3, 4$, the differences of $\mathrm{Re}\omega$'s are small. The dimensionless parameter $L$ is set to be unity for convenience. }
\label{fig:QNM_Q}
\end{figure}

Using the sixth-order WKB method, we calculate the QNMs with the overtone number $n=0$ and angular number $l=2$. Specifically, we plot QNMs with respect to $\hat{Q}$ in Fig.~\ref{fig:QNM_Q}. Again, the dimensionless rescaling introduced by Eqs.~(\ref{dimensionless parameter}), (\ref{lapsefunc2}), and (\ref{q1_2}) is adopted here.

$\mathrm{Re}\omega$ differs on a varying scale exponent. In general, it increases monotonically with the increasing of $\hat Q$ for a fixed $N$. For a large $N$, the value of $\mathrm{Re}\omega$ is apparently large for a fixed $\hat Q$, {\em e.g.}, $\mathrm{Re}\omega$ with $N=10$ is much larger than those with $N=0, 2, 3, 4$.
However, for a small $N$, such as $N=2, 3, 4$, $\mathrm{Re}\omega$ does not increase with an increasing $N$ monotonically when compared with $\mathrm{Re}\omega$ in the case of $N=0$. This result shows that the oscillation frequency of the CEPYM BHs with a small $N$ may be higher or lower than that of the EPYM BHs.

$\mathrm{Im}\omega$ also differs on a varying scale exponent, and its absolute value simply increases monotonically with an increasing $N$ for a fixed $\hat Q$. More significantly, the behavior of $\mathrm{Im}\omega$ provides the information about the second-order phase transition mentioned in Sect.~\ref{sec:2nd PT}.
Note that the relation between the dimensionless mass $m$ and dimensionless charge $\hat{Q}$ is  $\hat{Q}=m^{-2(D-2)/(D(D-3))}$. Substituting $m_{\rm Davies}$ into the relation, we obtain the corresponding values of $\hat{Q}_{\rm Davies} \approx 0.870$, $\hat{Q}_{\rm Davies} \approx0.447$, and $\hat{Q}_{\rm Davies} \approx 0.313$, in 4, 5, and 6 dimensions, respectively. In Fig.~\ref{fig:QNM_Q}, they are marked by the black vertical lines. We find that the Davies points correspond to the inflection points of the curves where the slopes begin to turn from negative to positive. The positions of these inflection points are irrelevant of the scale exponent, which shows that the second-order phase transition is irrelevant of the scale exponent as shown in Sect.~\ref{sec:2nd PT}. Consequently, the behavior of QNMs of the scalar perturbation proves from the point of view of BH dynamics that the second-order phase transition of the CEPYM BHs is independent of $N$. This relation between the dynamics and thermodynamics of the CEPYM BHs is presented
very clearly by the minimally coupled scalar field. If the non-minimally coupled scalar field,
for instance, the conformally coupled scalar field were considered, the QNMs would be
independent of the scale factor, and thus the inflection behavior of the QNMs with different
scale exponents would not be manifested.

In summary, we conclude that the scale factor has a significant influence on the dynamic properties of the CEPYM BH spacetimes.

\section{Conclusion}
\label{sec: conc}

In the present paper, we investigate the non-singularity of the CEPYM BH spacetimes, the thermodynamics and dynamics of the CEPYM BHs, and their relationship.

We prove that the CEPYM BH spacetimes are non-singular in two aspects. At first, the geometric quantities (the Ricci scalar  and Kretschmann scalar) are non-singular at the center of black holes. Secondly, the CEPYM BH spacetimes are geodesically complete.

We discuss the thermodynamics of the CEPYM BHs in the following three aspects.

(a) The Hawking temperature does not depend on the scale factor ${\cal S}(r)$, which is consistent with the result in Ref.~\cite{1605} where the Schwarzschild BH and its conformally related counterparts were analyzed. Meanwhile, the extreme configurations of the CEPYM BHs have a vanishing Hawking temperature, which implies no Hawking radiation.

(b) In the CEPYM BH spacetimes, $S_{D}=A_{D-2}/4$ no longer holds, because $S_D$ is independent of the scale factor $\mathcal{S}(r)$ while $A_{D-2}$ is dependent on $\mathcal{S}(r)$. As the Hawking temperature
and the BH mass are invariants of conformal transformations, the first law of BH thermodynamics maintains unchanged. As a result, the entropy of the CEPYM BHs is nothing else but the entropy of the EPYM BHs, that is, it is irrelevant to the scale factor ${\cal S}(r)$.

(c) The specific heat of the CEPYM BHs is independent of the scale factor ${\cal S}(r)$ and the Davies points of the CEPYM BHs associated with second-order phase transitions are same as those of the EPYM BHs. A CEPYM BH undergoes one second-order phase transition near its extreme configuration.

All the discussions about the Hawking temperature, the entropy, and the second-order phase transition lead to the conclusion that the thermodynamics of the CEPYM BHs is independent of the scale factor, which can be understood in the perspective of statistical physics and thermodynamics: A pure dilation of spacetime volume by ${\cal S}(r)$ should have no influence on the microscopic states of a BH, and thus no influence on its macroscopic thermodynamic properties.

For the dynamics of the CEPYM BHs, we obtain the real and imaginary parts of QNMs by using the sixth-order WKB method. The dependence of QNMs on the scale exponent is obvious.
${\rm Re}\omega$ increases monotonically with the increasing of $\hat Q$ for a fixed $N$. For a large $N$, for instance, $N\ge 10$, the value of $\mathrm{Re}\omega$ is apparently large. Quite interesting is that ${\rm Re}\omega$ with a small $N$, for instance, $N=2, 3, 4$, is not certainly greater than that with $N=0$ for a fixed $\hat Q$, which means that the oscillation frequency of the CEPYM BHs with a small $N$ may be higher or lower than that of the EPYM BHs.
Moreover, $|\mathrm{Im}\omega|$ increases monotonically with the increasing of the scale exponent for a fixrd $\hat Q$. The inflection points of $\mathrm{Im}\omega$ correspond to the Davies points which
are related to the second-order phase transition.
In other words, both the QNMs and the specific heat confirm the existence of second-order phase transitions. Therefore, the Hawking temperature Eq.~(\ref{HTsurface}) and specific heat Eq.~(\ref{SpecificHeat1}) are well-defined and indeed manifest the thermodynamic properties of the CEPYM BHs.

Finally, we conclude that the scale factor ${\cal S}(r)$ produces the distinct thermodynamic and dynamic effects in the CEPYM BH spacetimes: The thermodynamics does not depend on the scale factor ${\cal S}(r)$, while the dynamics does. This distinction reveals a new relationship between the thermodynamics and dynamics of the CEPYM BHs.
As the relation between the thermodynamics and dynamics of BHs plays~\cite{Hod} an important role in BH quantization, our result may provide helpful information in the study of BH quantization in conformal gravity.

\section*{Acknowlegement}

The authors would like to thank Y. Guo, C. Lan, and H. Yang for helpful discussions, in particular, to thank the anonymous referees for the helpful comments that improve this work greatly.
This work was supported in part by the National Natural Science Foundation of China under Grant No. 11675081.

\end{document}